# "Persistent" Insulator: Avoidance of Metallization at Megabar Pressures in Strongly Spin-Orbit-Coupled $Sr_2IrO_4$

Chunhua Chen[1,2], Yonghui Zhou[1]*, Xuliang Chen[1], Tao Han[3], Chao An[3], Ying Zhou[3], Yifang Yuan[1,2], Bowen Zhang[1,2], Shuyang Wang[1,2], Ranran Zhang[1], Lili Zhang[4], Changjing Zhang[1,3,5], Zhaorong Yang[1,3,5]*, Lance E. DeLong[6] and Gang Cao[7]*

[1] *Anhui Province Key Laboratory of Condensed Matter Physics at Extreme Conditions, High Magnetic Field Laboratory, Chinese Academy of Sciences, Hefei 230031, China*

[2] *University of Science and Technology of China, Hefei 230026, China*

[3] *Institutes of Physical Science and Information Technology, Anhui University, Hefei 230601, China*

[4] *Shanghai Synchrotron Radiation Facility, Shanghai Advanced Research Institute, Chinese Academy of Sciences, Shanghai 201204, China*

[5] *Collaborative Innovation Center of Advanced Microstructures, Nanjing University, Nanjing 210093, China*

[6] *Department of Physics and Astronomy, University of Kentucky, Lexington, KY 40506, USA*

[7] *Department of Physics, University of Colorado at Boulder, Boulder, CO 80309, USA*

*Corresponding authors: yhzhou@hmfl.ac.cn; zryang@issp.ac.cn; gang.cao@colorado.edu

**Abstract**

It is commonly anticipated that an insulating state collapses in favor of an emergent metallic state at high pressures as the unit cell shrinks and the electronic bandwidth broadens to fill the insulating energy band gap. Here we report a rare insulating state that persists up to at least 185 GPa in the antiferromagnetic iridate $Sr_2IrO_4$, which is the archetypical spin-orbit-driven $J_{eff} = 1/2$ insulator. This study shows the electrical resistance of single-crystal $Sr_2IrO_4$ initially decreases with applied pressure, reaches a minimum in the range, 32 - 38 GPa, then abruptly rises to fully recover the insulating state with further pressure increases up to 185 GPa. Our synchrotron x-ray diffraction and Raman scattering data show the onset of the rapid increase in resistance is accompanied by a structural phase transition from the native tetragonal *I*$4_1$/*acd* phase to an orthorhombic *Pbca* phase (with much reduced symmetry) at 40.6 GPa. The clear-cut correspondence of these two anomalies is key to understanding the stability of the insulating state at megabar pressures: Pressure-induced, severe structural distortions prevent the expected metallization, despite the 26% volume compression attained at the highest pressure accessed in this study. Moreover, the resistance of $Sr_2IrO_4$ remains stable while the applied pressure is tripled from 61 GPa to 185 GPa. These results suggest that a novel type of electronic Coulomb correlation compensates the anticipated band broadening in strongly spin-orbit-coupled materials at megabar pressures.

It is well established that a rare interplay of on-site Coulomb repulsion, U, and strong spin-orbit interactions (SOI) has unique, intriguing consequences in 4d- and 5d-transition metal oxides [1-15]. The SOI-driven $J_{eff}$ = ½ Mott insulating state in the 5d-transition metal oxide $Sr_2IrO_4$ is a profound manifestation of such an interplay [1, 2]. $Sr_2IrO_4$ adopts a canted antiferromagnetic (AFM) state [16] with a Néel temperature $T_N$ = 240 K [17-20] and an energy gap $\Delta \leq 0.62$ eV [21-23]. It exhibits key structural, electronic and magnetic features similar to those of $La_2CuO_4$, which has inspired expectations that novel superconductivity could emerge in $Sr_2IrO_4$ via electron doping [9]. However, there has been no experimental confirmation of superconductivity despite intensive experimental efforts [5].

It has become increasingly clear that the conspicuous absence of superconductivity in $Sr_2IrO_4$ is due in part to structural distortions; in particular, $IrO_6$ octahedral rotations play a crucial role in determining the ground state [5, 16, 24, 25]. The inherently strong SOI in $Sr_2IrO_4$ locks the canted Ir moments to the $IrO_6$ octahedra in a manner that is not seen in other materials, such as the cuprates [5, 16, 25, 26].

Despite the potential for broad and novel consequences of strong SOI in $Sr_2IrO_4$, the overwhelming balance of attention has been devoted to the possible existence of superconductivity [14]. The present work demonstrates a clear-cut, intriguing behavior of $Sr_2IrO_4$ that has thus far escaped notice: Early high-pressure studies [26, 27] indicated that $Sr_2IrO_4$ does not metallize up to 55 GPa, which sharply contrasts with the conventional view that a metallic state must either emerge or persist at high pressures as the unit cell volume shrinks and the electronic bandwidth broadens [28].

One of the most dramatic examples in support of traditional expectations is the recently discovered superconductivity in hydrogen sulfide above 200 K at megabar pressures [29]. These early studies also suggest that the *relative* strength of SOI with respect to the Coulomb correlation energy may diminish as pressure increases up to 90 GPa, thus, a superconducting state may occur in $Sr_2IrO_4$ [26].

Here we report electrical resistance, synchrotron x-ray diffraction and Raman scattering data for $Sr_2IrO_4$ at megabar pressures. The central finding of this study is that $Sr_2IrO_4$ remains as a robust insulator up to 185 GPa, in sharp contrast to the known behavior of most other materials. Lithium and sodium metals are noteworthy exceptions, as both metals become insulating near 40 GPa and 200 GPa, respectively, due to *p–d* hybridization of valence electrons and their repulsion by core electrons into the lattice interstices [30, 31]. The electrical resistance of $Sr_2IrO_4$ indeed initially decreases with pressure, reaching a minimum in the pressure interval, 32 - 38 GPa. However, the resistance behavior of $Sr_2IrO_4$ then takes a remarkable turn near 38 GPa, rapidly rising to fully recover the insulating state, which remains stable up to 185 GPa.

A structural phase transition from the native tetragonal *I*4$_1$/*acd* phase to a lower-symmetry orthorhombic *Pbca* phase occurs at a critical pressure 40.6 GPa, which is clearly associated with the rapid rise in the resistance. The pressure-induced orthorhombic *Pbca* phase is stabilized by both a rotation and tilt of $IrO_6$ octahedra. These structural distortions further weaken electron hopping, making $Sr_2IrO_4$ a "persistent" insulator. Perhaps our most remarkable observation is that the high electrical resistance (~$10^7$ Ω) of the sample remains essentially unchanged with a tripling of applied pressure, ranging from 61 GPa to 185 GPa, and despite a maximal, 26% volume compression. The direct correlations between structural, and the very unusual magnetic and electrical transport properties of $Sr_2IrO_4$ [24,25] highlights a critical role for the lattice degrees of freedom and the novel consequences of strong SOI. Further, our results shed light on the absence of a widely anticipated superconducting state in doped $Sr_2IrO_4$ [9], and point to the existence of non-traditional Coulomb correlation effects at high pressures in iridates.

The experimental details are listed in the Supplemental Material [32]. The measurements of the basal-plane electrical resistance, R, were conducted on a single crystal of $Sr_2IrO_4$ in two separate runs (Run1 and Run 2) that covered two pressure ranges of 0.6 - 54 GPa and 24 - 185.0 GPa, respectively.

We first focus on the basal-plane resistance R obtained in Run 1. For clarity, we plot the Run-1 data in two separate figures, namely, Fig. 1(a1) for 0.6 - 32.1 GPa, and Fig.1(a2) for 37.4 - 54.2 GPa. R increases monotonically with decreasing temperature at 0.6 GPa, in accordance with the insulating AFM state at ambient pressure [33]. With increasing pressure up to 32.1 GPa, R decreases by more than two orders of magnitude at low temperatures, but retains insulating behavior (Fig1(a1)). This trend reverses for P > 32 GPa, as shown in Fig.1(a2): R rises and almost fully recovers its initial value at ambient pressure. A corresponding T-P contour plot in Fig. 1(b) illustrates a U-shaped pressure dependence of R with a minimum near 32 GPa, which is consistent with previous results [26, 27].

We now turn to Run 2, which covers a wider and much higher pressure range of 24 -185.0 GPa, as shown in Fig. 1(c). This set of data confirms the results obtained in Run 1, although the lowest R occurs near 38 GPa, which is somewhat higher than 32 GPa observed in Run 1. (Given the inherent imperfections of high-pressure measurements, this difference is acceptable). More importantly, R remains essentially unchanged above 61 GPa; and an extended plateau in the pressure dependence of R between 61 and 185 GPa, as shown in Fig.1(d). Note that the low-temperature value of R is $\approx 10^7$ Ω at both ambient pressure and P > 61 GPa, suggesting that the insulating behavior observed under ambient conditions is fully recovered at higher pressures, in spite of expectations that the electronic structure should undergo significant changes with additional pressure increases $\approx$ 100 GPa!

The resiliency of the insulating state of $Sr_2IrO_4$ at such high pressures is extraordinary, and motivated a careful examination of the structural properties at high pressures. The results of *in situ* synchrotron XRD measurements at 300 K under pressures up to 74 GPa, are shown in Figure 2(a). Below 40.6 GPa, $Sr_2IrO_4$ retains the same structure as that at ambient pressure, i.e., the tetragonal space group *I*4$_1$/*acd*. As expected, all peaks progressively shift to higher angles, reflecting the shrinkage of the unit cell as P increases. An examination of the evolution of the diffraction peaks reveals a structural phase transition occurs at a critical pressure $P_c$ = 40.6 GPa. As seen from Fig. 2(b), the (112) peak intensity begins to become asymmetric at 40.6 GPa, and then gradually splits into two peaks upon further compression, signaling the occurrence of a structural transition. This is corroborated by the splitting of both (116) and (220) peaks, as well as the emergence of a new peak on the right shoulder of the (004) peak as the pressure reaches 51 GPa (see Figs. 2(c) and 2(d)). These emergent peaks become more pronounced with increasing P and are well indexed by an orthorhombic structure with space group *Pbca*, which requires both a rotation and tilt of the $IrO_6$ octahedra of $Sr_2IrO_4$. We note that a pressure-induced phase transition in $Sr_2IrO_4$ was reported in an early study; however, broadly overlapping peaks of XRD patterns at high pressures prevented a refinement of the pressure-induced space group [34].

It is clear that the crystal structure of $Sr_2IrO_4$ remains the ambient tetragonal phase below 37 GPa, and then assumes the orthorhombic *Pbca* phase at 40.6 GPa (Fig.3(c)). Further, when applied pressure is reduced from 74 GPa to 0.1 GPa, the ambient tetragonal phase is recovered, confirming the observed pressure-driven structural transition is intrinsic and reproducible (see the green line in Fig.2(a)). Indeed, the volume data are fitted well by the third-order Birch-Murnaghan equation of states [35]. The fitting results yield the ambient pressure volume $V_0 = 97.2\ (0.3)$, bulk modulus $B_0 = 218.2\ (10.4)$ and its first-order derivative $B_0' = 3.1\ (0.3)$ for the low-pressure tetragonal phase, and $V_0 = 81.3\ (0.6)$ Å$_3$, $B_0 = 340.0\ (17.2)$ GPa, and $B_0' = 13.1\ (0.2)$ for the high-pressure *Pbca* phase. Extrapolating pressure up to 185 GPa yields a 26% volume compression, compared to the ambient tetragonal phase. The standard Le Bail method is used for the structural refinement. A few representative refinement results at 26.9, 56.9, and 73.7 GPa are shown in Fig. S1 [32].

Raman scattering was employed as an effective and powerful tool for detecting small or local lattice distortions, as well as structural transitions. Figure 4(a) shows the selective Raman spectra at various pressures up to 65.6 GPa. At 1.0 GPa, there are four phonon peaks marked by $M_1$ (the merging of $A_{1g}$ and $B_{2g}$), $M_2$ ($A_{1g}$), $M_3$ ($B_{2g}$), and $M_4$ ($A_{1g}$), which are located at 177, 252, 388, and 561 cm$^{-1}$, respectively, in agreement with a previous report [34]. According to Ref. 34, the mode $M_1$ represents a rotation of the IrO$_6$ octahedra about the *c*-axis combined with a Sr displacement along the *c*-axis, while $M_2$ denotes a pure rotation of the IrO$_6$ octahedra about the *c*-axis. The mode $M_3$ is an in-plane bending of the IrO$_6$ octahedra, and $M_4$ is a stretching mode involving a modulation of the Ir-O (apical) distance. With increasing pressure, the Raman shift of all four modes first increases linearly, then shows clear slope changes near 15 GPa; this is particularly true for the $M_2$ mode that measures the octahedral rotation. Moreover, the $M_1$, $M_2$, and $M_3$ modes display distinctly abnormal red-shifts or phonon softening above 22.9 GPa, which gradually restore to blue-shift upon further compression. At and above the critical pressure $P_c$ = 40.6 GPa, a series of new peaks labeled by $P_{1-4}$ appear, which can be attributed to the structural transition detected in the XRD measurements. The Raman spectrum recovers its original form after decompression back to 3.6 GPa, suggesting sample integrity is preserved at the highest pressures (see the green line in Fig.4(a)).

The results of both XRD and Raman scattering provide a direct, crucial correlation of the lattice dynamics with the transport properties at high pressures, and demonstrate that the persistent insulating state at megabar pressures is related to the reduction in symmetry incurred in the transition from the *I*4$_1$/*acd* phase (32 symmetry group elements) to the much-lower-symmetry *Pbca* phase (8 symmetry group elements). This structural change involves not only rotations, but also titling of IrO$_6$ octahedra at P $\geq$ P$_c$. The striking stability of the resistance over such a broad pressure interval of 61 to 185 GPa suggests two competing forces are at work: 1) There is a tendency for band broadening that must accompany a 26% volume compression and favors metallic behavior. 2) There is a pressure-induced crystal distortion that generally weakens electron hopping and can lead to localization, which eventually prevails in the present case, given the fully recovered resistance for P > 61 GPa.

However, Sr$_2$IrO$_4$ defies simple Mott physics, in that the insulating state and long-range AFM order do not always accompany each other [5]. An early study indicates weak ferromagnetism vanishes near 18 GPa [26], which is in the vicinity of 15 GPa, where our Raman data clearly indicate a change in the IrO$_6$ rotation. The weak ferromagnetism is due to magnetic canting [16], which closely tracks the IrO$_6$ rotation [16, 25]. It is recognized that an elongation (compression) of the *c* axis weakens (enhances) the magnetic canting, or the weak ferromagnetism, and facilitates either a collinear AFM or a paramagnetic state [24]. Our XRD data show that the lattice *c/a* ratio increases significantly with rising pressure in both the tetragonal phase below P$_c$ (= 40.6) and the orthorhombic phase above P$_c$ (see Fig. 5). This certainly explains the disappearance of the weak ferromagnetism above 18 GPa, as reported previously [26]. However, it is not clear whether the canted AFM state evolves into a collinear AFM state or a paramagnetic state above 18 GPa. Nevertheless, the enhanced *c/a* ratio clearly indicates that Sr$_2$IrO$_4$ becomes more two-dimensional with increasing pressure, which is unfavorable for long-range magnetic order, and raises speculations that a spin liquid state may emerge at high pressures [36], which merits more investigations.

This work clearly documents a rare, persistent insulating state at megabar pressures, and its close correlation with a pressure-induced structural phase transition in $Sr_2IrO_4$. It clearly demonstrates the unique, crucial role the lattice symmetry and dynamics play in determining ground states in spin-orbit-coupled materials. Our results may offer some understanding of the conspicuous discrepancies between recent theoretical proposals and experimental results in iridates, including the question of superconductivity. More generally, the observed persistence of insulating behavior at megabar pressures raises an intriguing, fundamental issue: the strong exchange-correlation effects supported by a high density of states near the Fermi level are not effectively screened out by Fermi liquid interactions in $Sr_2IrO_4$, as traditionally anticipated from Mott physics. We speculate that when relevant orbitals are extended into highly directional bonds, Hartree-Fock mean-field theories, which treat breaking of spherical symmetry by electron-electron interactions via *spherical averaging* of self-consistent Coulomb fields, may not be well-suited for treating correlations and SOI in covalent "ligands" at very high electron densities.

**Acknowledgments**

We are grateful for the financial support from the National Key Research and Development Program of China (Grant No. 2018YFA0305700 and No. 2016YFA0401804), the National Natural Science Foundation of China (NSFC) (Grants

No. 11574323, No. U1632275, No. 11874362, No. 11804344, No. U1832209 and No. 11704387), the Users with Excellence Project of Hefei Science Center CAS (Grant No. 2018HSCUE012) and the Major Program of Development Foundation of Hefei Center for Physical Science and Technology (Grant No. 2018ZYFX002) The X-ray diffraction experiment was performed at the beamline BL15U1, Shanghai Synchrotron Radiation Facility (SSRF). GC acknowledges support from the US National Science Foundation via grants DMR 1712101 and 1903888. GC is thankful to Dr. Feng Ye, Dr. Bing Hu and Mr. Hengdi Zhao for useful discussions. LED research is supported by U.S. National Science Foundation Grant No. DMR-1506979.

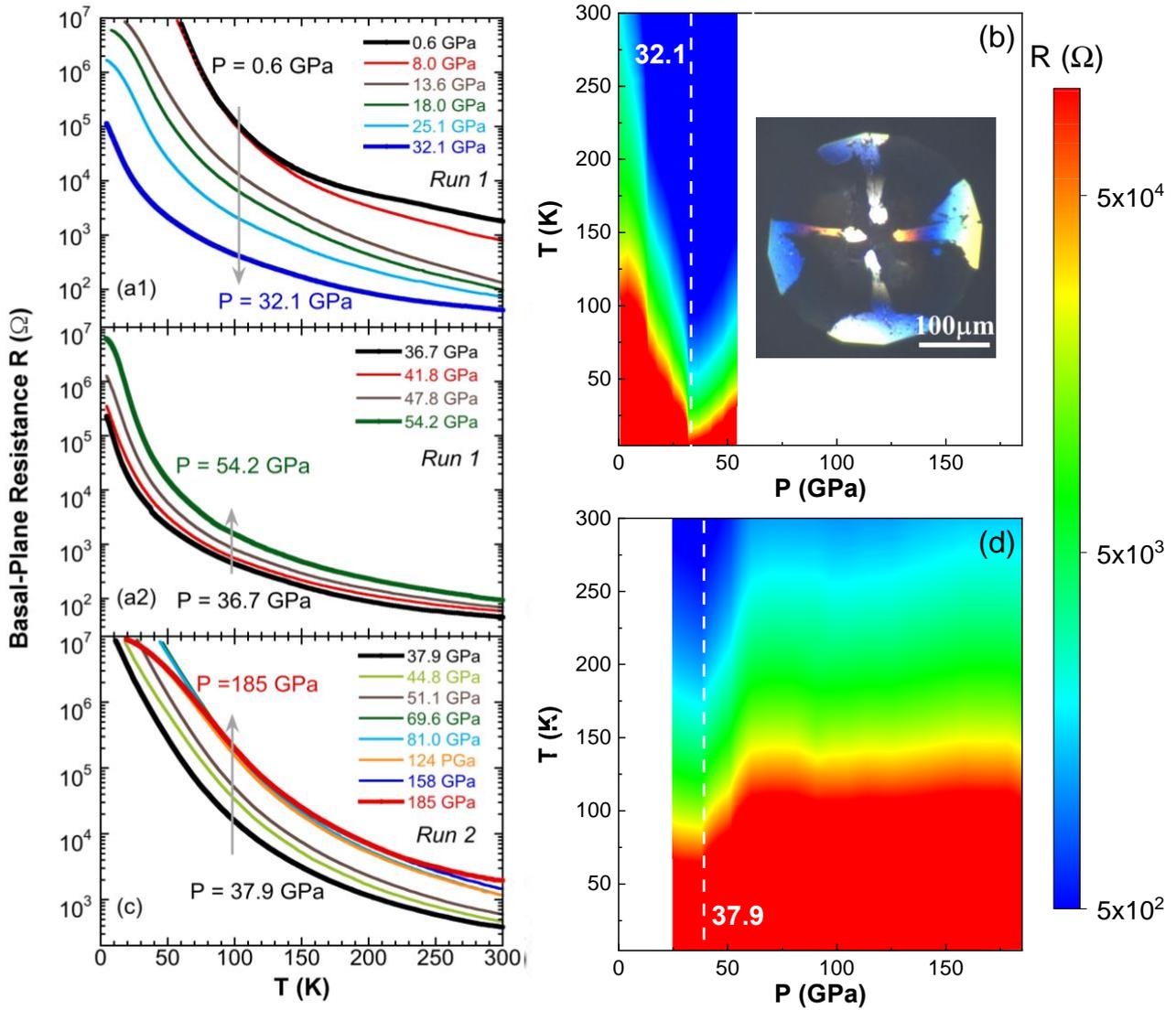

**Fig. 1**. The temperature dependence of the basal-plane resistance R over pressures ranging **(a1)** 0.6 – 32.1 GPa and **(a2)** 36.7-54.2 GPa for Run 1, and **(c)** 24.7 - 185 GPa for Run 2. Note the gray arrows that indicate the increase or decrease in R with increasing P. The corresponding contour plots **(b)** for the pressure range 0.6-54.2 GPa for Run 1 and **(d)** for the pressure range 24.7-185 GPa for Run 2. The colors red and blue represent the highest resistance and the lowest resistance R, respectively. The colors between them indicate intermediate resistance. The white dashed lines mark the pressure regime of 32.1 – 37.9 where R reaches its minimum. **Inset:** A snapshot of the diamond anvil cell with a sample at 27 GPa.

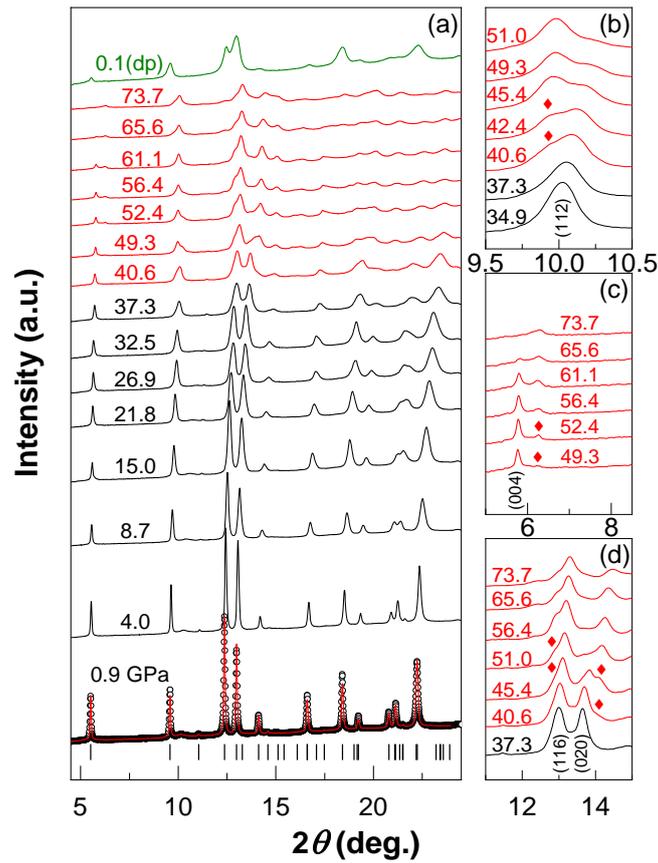

**Fig. 2. (a)** Representatives synchrotron x-ray diffraction patterns at room temperature spanning 0.9 to 73.7 GPa. The black curves represent the native $I4_1/acd$ phase (Z = 8), red patterns the pressure-induced $Pbca$ phase (Z = 4). The green line means the XRD curve after decompression. **(b, c, d)** The structural transition at 40.6 GPa is marked by red rhombi. The refinement results for 0.9 GPa are shown in (a). The experimental data are the solid circles, and the calculated data are red lines. Bragg peaks are represented by black vertical sticks.

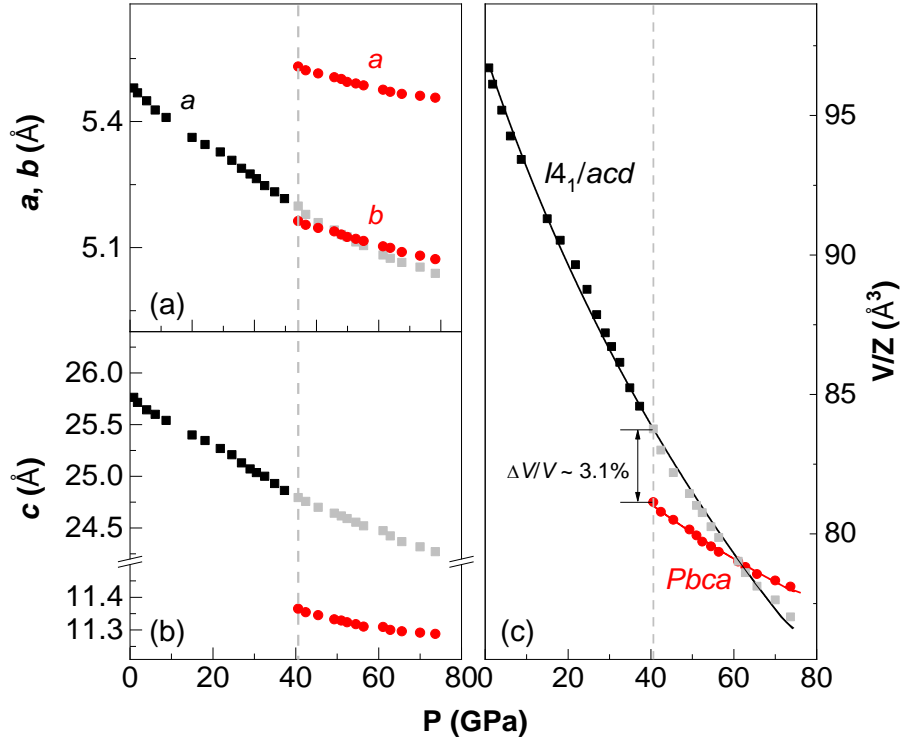

**Fig. 3**. Pressure dependence of the lattice parameters of **(a)** the *a* axis and the *b* axis, **(b)** the *c* axis, and **(c)** the unit cell volume. The black squares, red circles represent the lattice parameters of the native $I4_1/acd$ phase (Z = 8) and the pressure-induced *Pbca* phase (Z = 4), respectively. Note that the gray dashed lines mark the critical pressure $P_c$ = 40.6 GPa. For comparison and contrast, the lattice parameters for the native phase marked by the faint gray squares are plotted above $P_c$. The black and red solid lines represent the fitting for the phases with the Birch-Murnaghan equation of states.

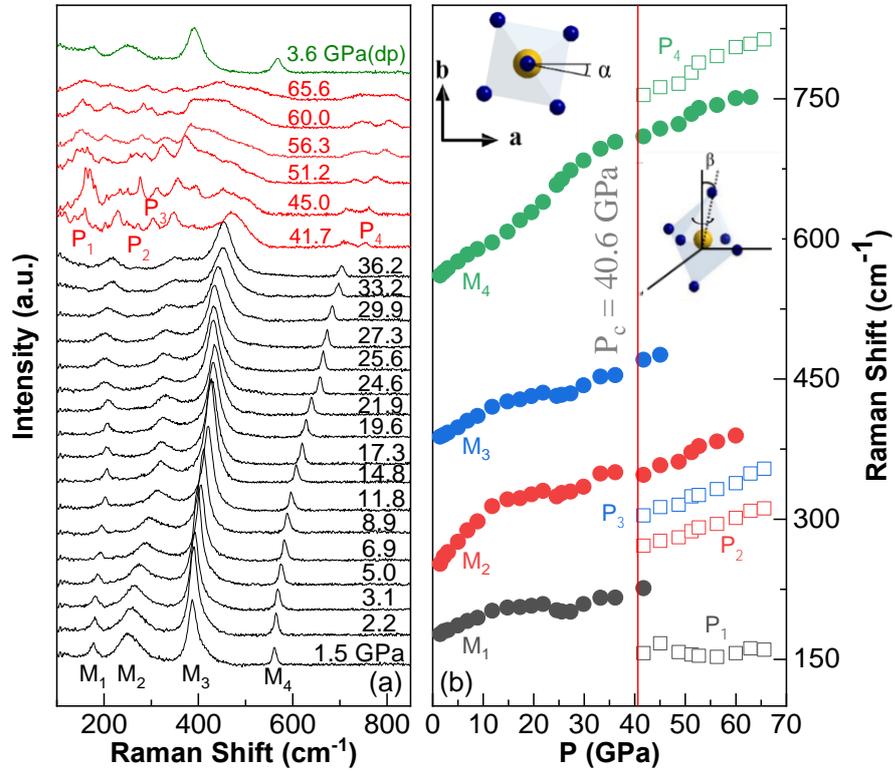

**Fig. 4. (a)** Selected room-temperature Raman spectra of $Sr_2IrO_4$ from 1.5 to 65.6 GPa. The black patterns include $M_{1-4}$; the red curves cover a series of new peaks marked $P_{1-4}$ after the transition. The uppermost green line is the spectrum after decompression. **(b)** Raman frequencies as a function of pressure for phonon modes. Note that the red solid line marks the critical pressure of 40.6 GPa, consistent with that obtained in the XRD measurements. **Insets:** The rotation of $IrO_6$ below $P_c$ and the rotation and tilt of $IrO_6$ above $P_c$.

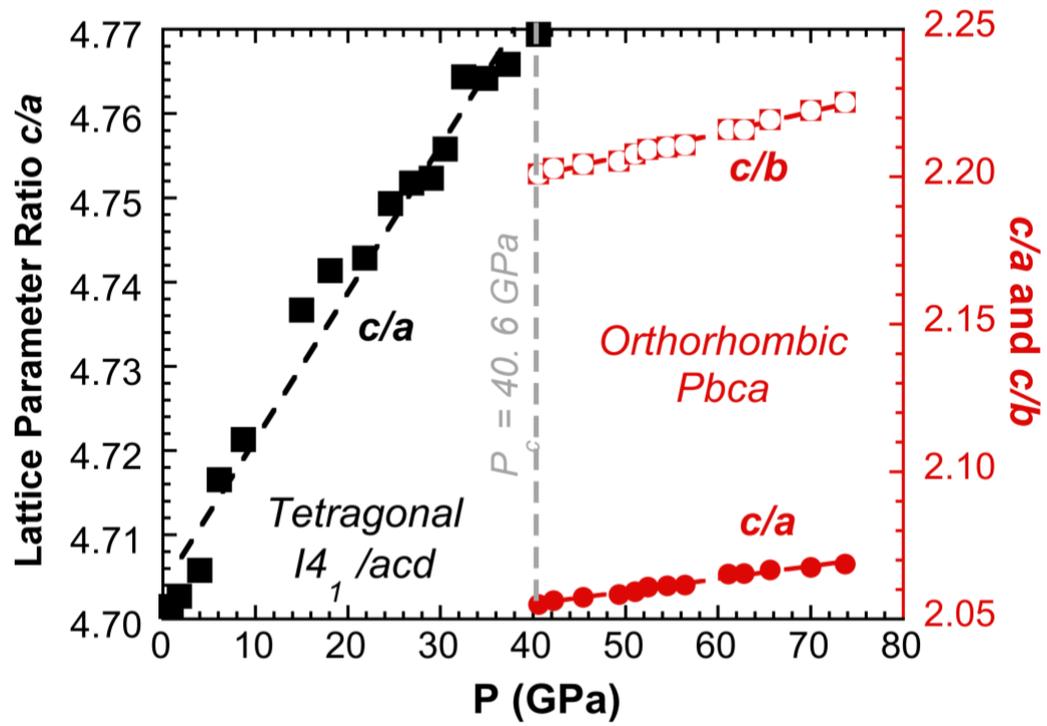

**Fig. 5.** The pressure dependence of the lattice parameter ratio of the *c* axis to the *a* axis, *c/a*, for the tetragonal phase (blue), and *c/a* and *c/b* for the orthorhombic phase (red).

*Supplemental Material*

# "Persistent" Insulator: Avoidance of Metallization at Megabar Pressures in Strongly Spin-Orbit-Coupled $Sr_2IrO_4$

Chunhua Chen[1,2], Yonghui Zhou[1]\*, Xuliang Chen[1], Tao Han[3], Chao An[3], Ying Zhou[3], Yifang Yuan[1,2], Bowen Zhang[1,2], Shuyang Wang[1,2], Ranran Zhang[1], Lili Zhang[4], Changjing Zhang[1,3,5], Zhaorong Yang[1,3,5]\*, Lance E. DeLong[6] and Gang Cao[7]\*

[1] *Anhui Province Key Laboratory of Condensed Matter Physics at Extreme Conditions, High Magnetic Field Laboratory, Chinese Academy of Sciences, Hefei 230031, China*
[2] *University of Science and Technology of China, Hefei 230026, China*
[3] *Institutes of Physical Science and Information Technology, Anhui University, Hefei 230601, China*
[4] *Shanghai Synchrotron Radiation Facility, Shanghai Advanced Research Institute, Chinese Academy of Sciences, Shanghai 201204, China*
[5] *Collaborative Innovation Center of Advanced Microstructures, Nanjing University, Nanjing 210093, China*
[6] *Department of Physics and Astronomy, University of Kentucky, Lexington, KY 40506, USA*
[7] *Department of Physics, University of Colorado at Boulder, Boulder, CO 80309, USA*

\*Corresponding authors: yhzhou@hmfl.ac.cn; zryang@issp.ac.cn; gang.cao@colorado.edu

$Sr_2IrO_4$ single crystals were grown by self-flux method described in reference [17]. Given the large magnitude of resistance, a two-probe method was employed to perform the high-pressure electrical transport measurements for a temperature range of 4.5-300 K in a Be-Cu diamond anvil cell (DAC) with a rhenium gasket. A freshly cleaved $Sr_2IrO_4$ single crystal was loaded with sodium chloride (NaCl) powder as the pressure transmitting medium and the electrical current was applied within the *ab* plane for Run 1. The sample was squeezed between the diamond anvil and insulation layer directly without pressure-transmitting medium in Run 2. The culet sizes of diamond were of 300 μm and 100 μm for Run 1 and Run 2, respectively. The Mao-Bell type symmetric DAC and the pressure transmitting medium Daphne 7373 were used for the measurements of high-pressure synchrotron powder x-ray diffraction (XRD) and Raman scattering; a pair of diamond with a culet of 300 μm diameter and rhenium gasket were also used for the measurements. The high-pressure XRD ($\lambda$ = 0.6199 Å) measurements were performed at room temperature using the beamline BL15U1 at the Shanghai Synchrotron Radiation Facility (SSRF). The DIOPTAS program [1] was employed for image integrations; the Le Bail method was used to fit the XRD patterns via the RIETICA program [2]. The Raman measurements were performed at room temperature on a freshly cleaved $Sr_2IrO_4$ single crystal using 633-nm He-Ne laser for excitation with the power below 10 mW to avoid sample damages and any heating effect. Pressure at room temperature was calibrated by the ruby fluorescence scale below 80 GPa [3] and the diamond Raman scale above 80 GPa [4], respectively.

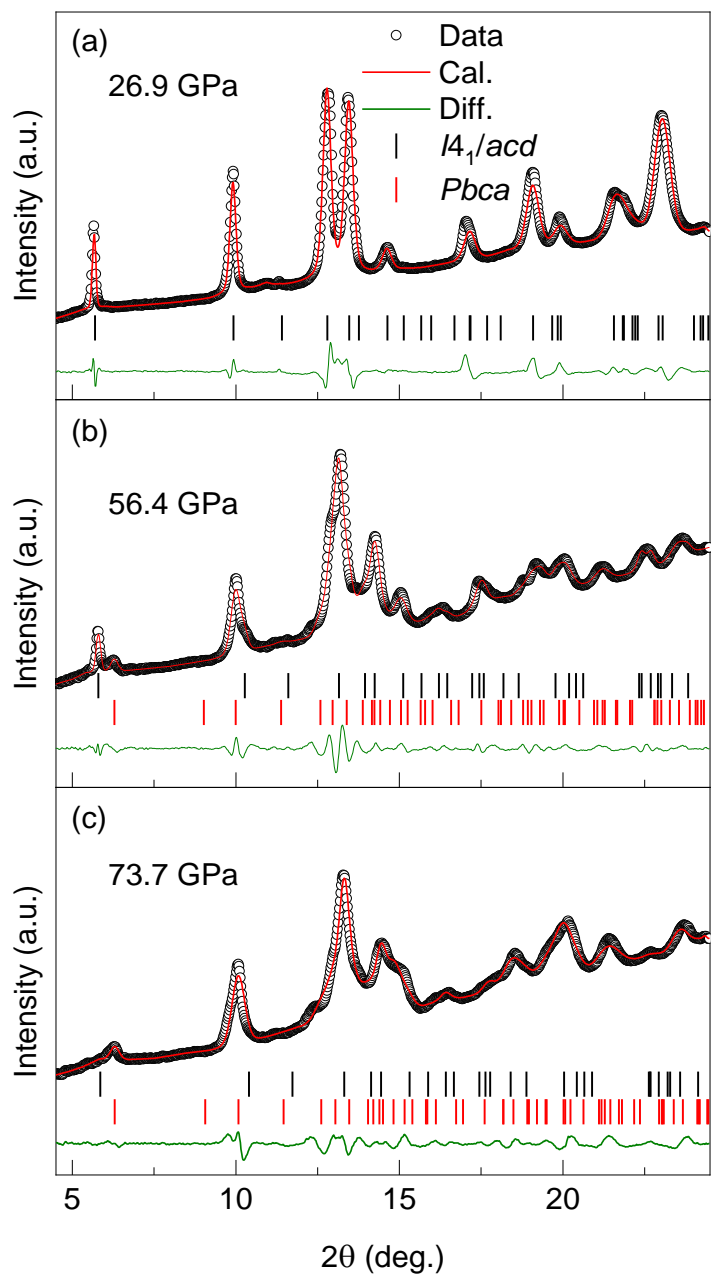

**Fig. S1**. The representatives of Le Bail fitting results of room-temperature XRD patterns at **(a)** 26.9, **(b)** 56.4 and **(c)** 73.7 GPa. The solid circles mark experimental data, calculated data are plotted as red lines, and the Bragg peaks are represented by vertical sticks (black, red for phase $I4_1/acd$, $Pbca$, respectively). The differences between experimental data and calculation are represented by dark green lines.